# Single-molecule imaging of DNA gyrase activity in living *Escherichia coli*

Mathew Stracy[1], Adam J.M. Wollman[2], Elzbieta Kaja[5], Jacek Gapinski[3], Ji-Eun Lee[2], Victoria A. Leek[4], Shannon J. McKie[4], Lesley A. Mitchenall[4], Anthony Maxwell[4], David J. Sherratt[1], Mark C. Leake[2,*], Pawel Zawadzki[1,3,*]

[1]Department of Biochemistry, University of Oxford, South Parks Road, Oxford, OX1 3QU, United Kingdom, [2]Biological Physical Sciences Institute (BPSI), Departments of Physics and Biology, University of York, York YO10 5DD, United Kingdom, [3]Molecular Biophysics Division, Faculty of Physics, A. Mickiewicz University, Umultowska 85, 61-614 Poznan, Poland, [4]Department of Biological Chemistry, John Innes Centre, Norwich Research Park, Norwich NR4 7UH, United Kingdom, [5]NanoBioMedical Centre, Adam Mickiewicz University, Umultowska 85, 61-614 Poznan, Poland.

*To whom correspondence should be addressed. Email: zawadzki@amu.edu.pl or mark.leake@york.ac.uk

The authors wish it to be known that, in their opinion, the first two authors should be regarded as joint First Authors.

# ABSTRACT

Bacterial DNA gyrase introduces negative supercoils into chromosomal DNA and relaxes positive supercoils introduced by replication and transiently by transcription. Removal of these positive supercoils is essential for replication fork progression and for the overall unlinking of the two duplex DNA strands, as well as for ongoing transcription. To address how gyrase copes with these topological challenges, we used high-speed single-molecule fluorescence imaging in live *Escherichia coli* cells. We demonstrate that at least 300 gyrase molecules are stably bound to the chromosome at any time, with ~12 enzymes enriched near each replication fork. Trapping of reaction intermediates with ciprofloxacin revealed complexes undergoing catalysis. Dwell times of ~2 s were observed for the dispersed gyrase molecules, which we propose maintain steady-state levels of negative supercoiling of the chromosome. In contrast, the dwell time of replisome-proximal molecules was ~8 s, consistent with these catalyzing processive positive supercoil relaxation in front of the progressing replisome.

## INTRODUCTION

The double-helical structure of DNA leads to major topological problems during DNA replication and transcription. As DNA and RNA polymerases translocate along the chromosome they cause local over-winding of DNA ahead of them; if excessive positive (+) supercoiling accumulates it can inhibit the progress of the enzymes, leading to a shutdown of these essential cell processes. Furthermore, (+) supercoiling, which accumulates ahead of the replication fork, can diffuse backwards causing entanglement of daughter chromosomes, which must be unlinked before cell division can occur. In *Escherichia coli* these topological problems are resolved by two type II topoisomerases, DNA gyrase and DNA topoisomerase (topo) IV, which are essential enzymes that change topology by introducing transient double-stranded breaks into DNA and pass a second double-stranded DNA segment through the break before resealing it (1) (Figure 1A). Gyrase, the focus of this study, is formed from a dimer of GyrA, primarily responsible for DNA binding, and two GyrB subunits, which provide the ATPase activity.

In *E. coli*, the chromosome is maintained in a negatively (-) supercoiled state, and the appropriate level of supercoiling is important for regulation of almost all processes which take place on DNA, including transcription, replication, repair and recombination (2,3). For example, the expression level of many genes, including gyrase itself, is regulated by the level of supercoiling (4). Gyrase is unique in its ability to introduce (-) supercoils into DNA, and is therefore the central enzyme responsible for maintaining supercoiling homeostasis (5-8); however, local DNA supercoiling is constantly being altered by ongoing replication, transcription and repair. The activities of gyrase must therefore be responsive to these processes taking place in different regions of the chromosome (9,10).

The most acute topological problem arises during DNA replication, which is performed by two replisomes traveling in opposite directions around the circular chromosome at speeds of up to 1000 base pairs per second (bp/s) (11,12). Without the action of type II topoisomerases, replication of the 4.6 Mbp E. coli chromosome would result in two daughter chromosomes interlinked with a linking number of more than 440000 (given the DNA helical repeat of 10.4 base pairs). Type II topoisomerases change the linking number by 2 each catalytic cycle, and must therefore perform over 220 000 catalytic events before segregation can occur. When the replisome is prevented from rotating around the DNA helix as it progresses, as originally suggested by Liu & Wang (13), (+) supercoils rapidly accumulate ahead of the replication fork. On the other hand, any rotation of the replication fork (14) allows (+) supercoils ahead of the fork to diffuse backwards forming precatenanes between the newly-replicated daughter chromosomes, which must be unlinked prior to chromosome segregation. Gyrase is inefficient in decatenation, and is believed to act ahead of the fork relaxing (+) supercoils, whereas topo IV acts preferentially behind the fork removing precatenanes (11,15,16).

To allow the replisome to maintain its incredibly high translocation rate, the two type II topoisomerases must relax up to 100 (+) supercoils per second for each fork (assuming a replisome translocation rate of 1000 bp/s, and DNA helical repeat of ~10bp) (Figure 1B). *In vitro,* the catalytic cycle for both gyrase and topo IV has been measured at ~2 s, with each cycle removing 2 supercoils (17-19), suggesting that up to 100 enzymes would be required per fork to keep up with the replication rate in live bacteria. Early studies of chromosome fragmentation in *E. coli* cells using the gyrase targeting drug, oxolinic acid (20), suggested that gyrase may be clustered near the replication fork. However, this raises the question of how so many gyrase enzymes can be acting ahead of the replication fork, while avoiding extremely toxic collisions with replication machinery. In single-molecule magnetic tweezers experiments *E. coli* gyrase was shown to act processively (18), confirming previous ensemble observations (1) and demonstrating that it is capable of performing multiple catalytic events without dissociating from DNA while relaxing (+) supercoils and introducing (-) supercoils (Figure 1A). More recent *in vitro* experiments on *Bacillus anthracis* gyrase suggests that gyrase 'bursting' activity might relax high levels of (+) supercoiling at faster rates (19). It remains to be established whether gyrase behaves processively or not *in vivo*, and whether its catalytic mode depends on the local supercoiling environment.

The action of gyrase is also essential for unperturbed transcription. Since coupling between RNA polymerase (RNAP) and (poly)ribosomes inhibits rotation of the transcription machinery, (+) supercoils accumulate ahead, and (-) supercoils behind, elongating RNAPs (Figure 1C) (7,9,21). While the rate of introduction of supercoils by a single RNAP is slow compared to replication (~60 bp/s, or ~6 (+) supercoils/s) (7-9), it is far more abundant. In a cell with 2 replisomes there are up to 2000 RNAPs (22), introducing more (+) supercoiling overall than replication, but distributed throughout the chromosome instead of accumulated in one region. The relative contribution of transcription and replication to gyrase activity is not clear.

We aimed to understand how gyrase acts in live *E. coli* cells and how topological problems arising during replication and transcription are resolved. Live cell epifluorescence showed that gyrase forms foci colocalized with active replication forks. However, single-molecule Slimfield (23,24) and photoactivated-localization microscopy (PALM) (25), showed that replication-dependent gyrase clusters comprise ~12 enzymes per replisome, while the remaining ~300 functional immobile enzymes interacted with the chromosome elsewhere to maintain steady-state levels of (-) supercoiling. An additional ~300 enzymes transiently interacted with dispersed regions of the chromosome. Measuring the dwell time of gyrase bound to DNA revealed that most gyrase remain immobile for ~2 s, whereas enzymes in the vicinity of the replisome had a ~8 s dwell time, suggesting that when an excessive (+) supercoiling is present due to the fast progression of the fork, gyrase performs multiple rounds of catalysis without dissociating from DNA.

## MATERIALS AND METHODS

### Bacterial strains

All strains were derivatives of Escherichia coli K-12 AB1157 (26). Replacement of endogenous genes with C-terminal fluorescent fusions was performed using λ-Red recombination with an frt-flanked kanamycin resistance (kan) cassette (27) using the primers listed in Table S1. The stains used in this study are: GyrApam (*gyrA::PAmCherry kan*); GyrBpam (*gyrB::PAmCherry kan*); PZ291 (*gyrA::mYPet kan*); PZ171 (*gyrA::PAmCherry kan, mYPet::DnaN frt*); PZ223 (*gyrA::mYPet kan, mCherry::DnaN frt*). See the Supplementary Materials and Methods for complete details of strain construction.

### Sample preparation

Strains were streaked onto LB plates containing the appropriate antibiotics. Single colonies were inoculated into M9 media supplemented with 0.2% glycerol and grown overnight at 37°C to $A_{600}$ 0.4-0.6, diluted into fresh M9 glycerol and grown to $A_{600}$ 0.1. Cells were centrifuged and immobilized for imaging on 1% agarose (Bio-Rad) pads (made by mixing low-fluorescence 2% agarose in dH$_2$O 1:1 with 2x growth medium) between 2 glass coverslips (Supplementary Methods).

**Epifluorescence and colocalization microscopy**

Wide-field epifluorescence was performed using an Eclipse TE2000-U microscope (Nikon), equipped with 100x/NA1.4 oil objective and a Cool-Snap HQ$^2$ CCD. For colocalization analysis cell outlines were defined from phase contrast images using MicrobeTracker software (28). The positions of foci formed by mCherry-DnaN were established with Gaussian fitting (Supplementary Methods). Pairwise distances between the center of the brightest GyrA-mYPet pixel and the centroid of the nearest DnaN localization were calculated in MATLAB (MathWorks) from the square root of the squares of the summed coordinates in x and y. To determine the distribution of distances expected from chance GyrA localizations we calculated distances between a pixel randomly positioned within the cell and the centroid of the nearest DnaN focus. A threshold of 2 pixels (256 nm) was chosen to define colocalization.

**Photoactivated Localization microscopy**

PALM microscopy was performed using a custom-built single-molecule microscope described in the Supplementary Materials and Methods. Photoactivatable mCherry activation was controlled with a 405 nm wavelength laser, and the photoactivated fluorophores were imaged with a 561 nm laser at 15.48 ms/frame for 30,000 frames. Data analysis was performed in MATLAB (MathWorks). Fluorescent signals from individual PAmCherry molecules in each frame were localized to ~40-nm precision by elliptical Gaussian fitting. Brightfield cell images were recorded from an LED source and condenser (ASI Imaging), and cell outlines were segmented with MicrobeTracker software (28). For colocalization analysis of super-resolved gyrase localizations with the replisome, snapshots of mYPet were taken with 488 nm excitation prior to PALM imaging of PAmCherry.

**Single-particle tracking and diffusion analysis**

Localizations from PALM movies were linked together into trajectories using a MATLAB implementation of the algorithm described in ref (29). Positions were linked to a track if they appeared in consecutive frames within a window of 5 pixels (0.48 µm). In rare cases when multiple localizations fell within the tracking radius, tracks were linked such that the sum of step distances was minimized. We distinguished DNA-bound and diffusing proteins by calculating an apparent diffusion coefficient $D^*$=MSD/(4Δ$t$) from the mean-squared displacement (MSD) for each track with at least 4 steps at Δ$t$=15ms (30). Immobile molecules have a non-zero $D^*$ value due to the localization uncertainty in each measurement, $\sigma_{loc}$ (40nm), which manifests as a positive offset in $D^*$ of ~0.1µm$^2$s$^{-1}$. Errors in $D^*$ and fractions are SEM from fitting to at least 4 independent experimental repeats. Significance testing was performed using 2-sample *t*-tests of the fraction of immobile molecules extracted from these fits (Supplementary Materials and Methods).

**Dwell-time distributions using long exposure times** Long duration GyrA-PAmCherry binding was recorded at low continuous 561 nm excitation intensities using 1 s exposure times. The probability of observing a particular on-time is the product of the binding time and bleaching probabilities (30). The bleaching time distributions were measured independently using a control protein, MukB-PAmCherry, whose dwell time was previously shown to be ~1 min >> bleaching time (31) . MukB-PAmCherry was imaged with the same imaging conditions. On-time and bleaching time distributions were fitted with single-exponential functions to extract exponential time constants $t_{on}$ and $t_{bleach}$, and the binding time constant calculated as $t_{bound} = t_{on}*t_{bleach}/(t_{bleach}–t_{on})$. To determine binding times near the fork snapshots of mYPet-DnaN were taken prior to PALM imaging. DnaN foci were localized with Gaussian fitting and GyrA tracks within 200nm of a focus were used for binding time analysis. The bleaching time, $tbleach$ = 1.16 ± 0.04. The uncorrected $t_{on}$ time constants from 7 experimental repeats are shown in Table S2.

**Slimfield microscopy**

Slimfield microscopy was performed on a dual-color custom-made laser excitation single-molecule fluorescence microscope which utilized narrow epifluorescence excitation of 10 µm full width at half maximum (FWHM) in the sample plane to generate Slimfield illumination from a 514 nm 20mW laser passed through a ~3x Keplerian beam de-expander. Illumination was directed onto a sample mounted on an xyz nanostage (Mad City Labs, the Dane County, Wisconsin, USA). Imaging was via a custom-made color splitter utilizing a dual-pass green/red dichroic mirror centered at long-pass wavelength 560 nm and emission filters with 25 nm bandwidths centered at 542 nm and 594 nm (Chroma Technology Corp., Rockingham, Vermont, USA) onto an Andor iXon 128 emCCD camera, magnified to 80 nm/pixel.

For dual color imaging we acquired 10 frames of brightfield, defocused to image the cell boundary, then acquired mCherry images by exciting with 1 mW 561 nm laser until bleached after 500 frames. Then, the mYPet images were acquired, exciting with 10 mW of 514 nm laser for 500 frames. Brightfield imaging was performed with zero gain at 100 ms exposure time while single-molecule fluorescence was performed at maximum gain at 5ms/frame, with the addition of the 561 nm laser for mCherry. Imaging of the single label mYPet-GyrA strain utilized only 514 nm laser excitation.

Stoichiometry was determined using a method which relies of step-wise photobleaching of fluorescent protein checked against surface immobilized purified mYPet using Chung-Kennedy filtration on single-molecule intensity bleach traces(24,32-38). Probability distributions for the relative displacement of GyrA-DnaN foci and for the stoichiometry of GyrA foci were rendered using kernel density estimation (KDE), a convolution of the data with a Gaussian kernel which has an advantage in objectifying the appearance of the distribution as opposed to using semi-arbitrary bin widths on a histogram plot. The kernel width was set to the appropriate experimental precision (0.7 molecules for the stoichiometry distribution and 40 nm for the distance estimates). See Supplementary Materials and Methods.

**RESULTS**

**Gyrase foci colocalize with the replisome**

To characterize gyrase activity in live cells we replaced the endogenous *gyrA* gene with a fusion to the fluorescent protein mYPet. Cells with *gyrA-mYPet* showed normal growth indicating the fusion is functional (Supplementary Figure S1A), and purified GyrA-mYPet showed normal supercoiling activity *in vitro* (Figure 1D). Using epifluorescence, gyrase formed foci in 70 ± 6 % (±SD) of cells, with the remaining cells showing a diffuse fluorescent signal, consistent with gyrase localization throughout the chromosome (Figure 2).

Since gyrase is thought to remove (+) supercoils ahead of the replication fork, we constructed a strain expressing GyrA-mYPet and a replisome marker mCherry-DnaN (11). We find that the region with highest gyrase density is frequently colocalized with the replisome (Figure 2A), reflecting earlier findings from *B. subtilis* (39). To quantify colocalization we used Gaussian fitting to localize the replisome foci and examined the cumulative distributions of distances between the brightest pixel of gyrase signal and the nearest replisome focus within each cell (Figure 2B). To control for colocalization due to random coincidence we performed the same analysis with a simulated random gyrase focus position within the same cells, showing that 80±4% of the brightest gyrase pixels were located within 2 pixels (256 nm) from the replisome, compared to 15 ± 3% from random coincidence.

In the slow growth conditions used for our experiments, a single round of replication takes only ~2/3 of the cell doubling time, leaving a population of young cells that have not initiated replication or cells approaching division that have completed replication (Figure 2A). Since the fraction of cells lacking replication foci (~25% identified with spotFinder (28)) was similar to the fraction of cells lacking gyrase foci (~30%), we asked whether the presence of gyrase foci was dependent on ongoing replication; in cells without a DnaN focus only 30 ± 10% of these non-replicating cells had a distinct gyrase focus. Taken together, this analysis suggests that distinct gyrase foci are largely associated with replication forks.

**Slimfield microscopy reveals gyrase clusters of ~12 enzymes**

Epifluorescence microscopy provides a description of the ensemble behavior of fluorescently labeled proteins inside cells, however it cannot provide a quantitative assessment of protein activity at the level of individual molecules. To enable single-molecule quantification of gyrase localization we used Slimfield microscopy on GyrA-mYPet in live cell (23,40), providing a ~40 nm spatial precision over a millisecond temporal resolution to enable blur-free analysis of individual proteins (SI Movie 1). Qualitatively, the patterns of GyrA localization with respect to DnaN (Figure 3A, Supplementary Figure S3) were similar to those observed earlier for epifluorescence (Figure 2A). Using analysis based on the integrated pixel intensity of Slimfield images (40) we quantified the GyrA copy number, giving 1300-3300 molecules per cell across all cells, which agrees broadly with earlier estimates based on immuno-gold electron microscopy of fixed *E. coli* cells (41).

To estimate the number of gyrase in localized clusters we used custom-written localization software to automatically track GyrA foci (42). We determined the stoichiometry of each as the initial focus brightness divided by the brightness of a single mYPet (32) (Material and Methods). Given the rate of relaxation of 2 positive supercoils per ~2 s previously reported for gyrase (18,43-45) and assuming minimal involvement of topo IV, we expected clusters to comprise of up to 100 gyrase (since 100 enzymes are required to keep up with a replication rate of 1000 bp/s). However, the intensity of these foci indicated a mean of 24±2 (±SEM) GyrA molecules (i.e. just 12±1 putative heterotetramer enzymes); note a key advantage of this single-molecule approach over ensemble methods is to render not just the mean value but also the full probability distribution, which we measure as having a broad range from a minimum of 2 molecules to over 100 per focus (Figure 3B). Using numerical integration of the overlap integral between green and red channel foci we observed that ~85% of all foci were colocalized with DnaN, comparable to epifluorescence. The relative separation between DnaN and GyrA foci centers was not peaked at zero but instead had a mean of 135±14 (±SEM) nm, exhibiting a unimodal distribution which extended up to ~400 nm (Figure 3C), larger than the ~50 nm replisome diameter, suggesting that gyrase does not act in tight proximity to the replisome. The hypothesis that gyrase acts at a distance from the fork might explain how collisions between the replisome and gyrase performing catalysis are prevented, however we note that while DnaN forms diffraction-limited foci, it has been shown that their dissociation rate is slow and hence the focus centroid may be slightly behind the replication fork (46).

**Photoactivated-localization microscopy and single-particle tracking of gyrase**

To explore the mobility of single gyrase we used photoactivated-localization microscopy (PALM), combined with single-particle tracking (sptPALM) (25), enabling localization and tracking of individual GyrA by controlling the photoactivation of a photoactivable fluorescent protein such that on average one fluorophore was active per cell at any given time. We labeled GyrA genomically with photoactivable mCherry (PAmCherry) (Figure 1D, Supplementary Figure S1A) and imaged cells with a PALM microscope at 15 ms intervals for 30,000 frames. Linking consecutive GyrA localizations from each frame into tracks allowed us to track gyrase movement until photobleaching (Figure 4B) (25,30).

We calculated an apparent diffusion coefficient ($D^*$) for each GyrA from the mean squared displacement of its track (Materials and Methods). We fitted an analytical expression (22,47) to the distribution of $D^*$ values from all 85529 measured tracks. We found that the distribution of $D^*$ values was best described by a three-species model: immobile (46 ± 5%; $D_{imm}$=0.1 $\mu m^2 s^{-1}$ set by the localization precision), slow-diffusing (42 ± 4%; $D_{slow}$=0.25± 0.01 $\mu m^2 s^{-1}$) and fast-diffusing (12 ± 4%; $D_{fast}$=0.82 ± 0.10 $\mu m^2 s^{-1}$) populations (Figure 4C). Fitting one or two species to the $D^*$ distribution provided a poor description of the data (Supplementary Figure S1B,C).

We interpret immobile tracks as DNA-bound gyrase and fast-diffusing tracks as gyrase undergoing free 3D diffusion, possibly GyrA molecules not incorporated into functional gyrase heterotetramers with GyrB. Slow-diffusing gyrases have lower mobility than expected for free 3D diffusion, consistent with transient interactions with DNA without engaging in stable binding required for catalysis.

To asses gyrase expression, we photoactivated and tracked all GyrA-PAmCherry molecules present in each cell, indicating a mean of ~1450 ± 550 (SD) GyrA per cell (Figure 4A). We note that the copy number measured using PALM may underestimate the true copy number due to a population of PAmCherry which do not become fully photoactivatable (although this has never been characterized in bacteria) (48). Nevertheless, this estimate falls within the range estimated earlier from Slimfield microscopy, which does not use photoactivatable fluorescent proteins and hence does not suffer the same technical issue. For simplicity we have based all calculations derived from PALM experiments on the unmodified mean copy number of 1450 GyrA, but we acknowledge that the true copy number could potentially be up to two-fold larger.

To estimate the proportion of GyrA able to form functional heterotetramers, we treated GyrA-PAmCherry cells with ciprofloxacin, which traps gyrase on DNA by stabilizing the covalently linked DNA-gyrase complex formed during catalysis (49). We find that 80±3% of GyrA are immobile after drug treatment (Figure 4D), a significant increase (p = $6x10^{-5}$) from unperturbed cells and more than twenty-fold higher than early estimates of ~45 stabilized gyrase based on chromosome fragmentation with the much less potent quinolone, oxolinic acid (50). Since ciprofloxacin is not known to be able to capture gyrase subunits not incorporated into heterotetramers, and only stabilizes enzymes during catalysis, this demonstrates that the GyrA-PAmCherry stabilized on DNA after ciprofloxacin treatment were incorporated into functional enzymes that underwent catalysis. Assuming a copy number of 1450 GyrA subunits, of which 12% are fast-diffusing putative unincorporated subunits, our findings show that in an average cell there is enough GyrA to form ~600 functional enzymes, of which ~300 are DNA-bound and likely performing catalysis.

**Gyrase activity in cells not undergoing replication or transcription**

Epifluorescence microscopy indicates that gyrase foci are less common in cells not undergoing replication (Figure 2C). These cells show only a minimal reduction in the fraction of DNA-bound, immobile GyrA compared to replicating cells (Figure 4C and 5A) from 46 ± 5% immobile GyrA to 44 ± 5%, within statistical error, equating to a difference of just ~15 additional gyrase enzymes per cell (with 2 replisomes), broadly consistent with Slimfield observations suggesting an average of ~12 gyrase associated with each replisome.

We constructed a mYPet-DnaN, GyrA-PAmCherry strain to determine positions of replisomes relative to PALM-tracked gyrase (Figure 5B). The fraction of immobile gyrase 'proximal' (within 200 nm) to the replisome is 16 ± 12% which, when corrected by a fraction of simulated randomly distributed gyrase in the same region (8 ± 0.5%), equates to ~25 more gyrase located next to both replisomes than expected from a random distribution, consistent with the small reduction of immobile gyrase observed in non-replicating cells (Figure 5A). In summary, on average only 8-12 gyrase are involved in relaxation of (+) supercoiling introduced by each replisome, and most of the remaining ~300 DNA-bound gyrases are immobile throughout the rest of the chromosome. To test where immobile gyrase is catalytically active we treated cells with ciprofloxacin and analyzed the distribution of immobile molecules within the cells. We found immobile gyrase throughout the chromosome (Supplementary Figure S2B), suggesting that molecules close to and far from the replisome perform catalysis.

Gyrase not associated with the replisome could be relaxing (+) supercoils introduced by RNAP or be involved in maintaining steady-state levels of chromosomal (-) supercoiling. To distinguish these possibilities, we treated cells with the transcription initiation inhibitor rifampicin, resulting in a moderate reduction (by 11%) in the fraction of immobile gyrase (Figure 5C), consistent with earlier experiments which showed that rifampicin reduces plasmid supercoiling (51) Nevertheless, since 33% of gyrase remain immobile after rifampicin treatment, this suggests that gyrase performs its activity even when no (+) supercoils are being introduced due to transcription. We conclude that the role of the majority of gyrase in the cell is not directed towards relaxing (+) supercoiling introduced by replication, but rather towards maintaining steady-state chromosomal supercoiling.

**Different modes of gyrase**

To address the conundrum of how a low number of gyrase in the vicinity of the replisome can relax up to 100 supercoils per second, we aimed to determine whether the catalytic mode depended on proximity to the replisome. To do this we measured the binding time of gyrase inside live cells using sparse photoactivation with a low excitation intensity and long (1s) exposure time. Under these conditions mobile gyrases are motion blurred, whereas immobile molecules appear as distinct diffraction-limited foci (30,52) (Figure 6A).

The observed dwell time for gyrase was corrected for photobleaching as described previously (30), giving a mean binding time of 2.4 ± 0.5s (Figure 6B, Figure S4). As a control we also measured the binding time of topo IV using a ParC-PAmCherry fusion strain from our previous study, described in reference 16. For topo IV we measured a similar binding time (1.7 ± 0.2 s), consistent with the rate of ATP hydroysis estimated *in vitro* for both enzymes (18,53). Ciprofloxacin resulted in a drastic increase in the fraction of immobile molecules (Figure 4D) as well as increasing the binding time (to at least 30 s, the upper limit of our assay), indicative of gyrase trapped during its catalytic cycle (Figure 6B). We suggest that bound gyrase exhibiting binding times of ~2.5 s are undergoing single rounds of catalytic activity, however we cannot exclude the possiblity that some gyrase bind DNA without performing catalysis.

While the observed binding time for gyrase is consistent with rates measured *in vitro* (17,18), it does not resolve the puzzle of how gyrase foci comprised of only ~10 molecules relax (+) supercoils at a rate sufficient for replication fork progression at up to 1000 bp/s. By taking a snapshot of replication foci prior to measuring binding times, we categorized binding events taking place within ('proximal') or beyond 200 nm ('distal') from a mYPet-DnaN replisome marker. The binding time of distal gyrase (2.5 ± 0.4 s) shows no significant difference from 2.4 ± 0.5 s measured for the entire population (Figure 6C); however, proximal gyrase has a significantly longer binding time (7.7 ± 1.5 s). We propose that the longer binding time close to the replisome results from gyrase performing multiple rounds of catalytic activity without dissociating, which is facilitated by the high level of (+) supercoiling ahead of the fork.

**DISCUSSION**

DNA gyrase has been the subject of many biochemical and structural studies since its discovery in 1976 (1,43), however, many questions remain regarding how it acts in living cells. For example, *in vitro* gyrase can relax (+) supercoils, and also introduce (-) supercoils into relaxed DNA. Yet, little is known about what proportion of gyrase activity is directed towards different DNA substrates in the cell: removing (+) supercoiling introduced by replication, removing (+) supercoiling introduced by transcription, and maintaining steady-state (-) supercoiling of the chromosome. The relative activities of gyrase and topo IV during replication also remains a mystery. Furthermore, while *in vitro* studies have observed different modes of gyrase catalysis, it remains to be established if the catalytic mode depends on the substrate *in vivo*. In this work we have used a combination of live-cell fluorescence microscopy techniques, with the aim of bridging the gap between our understanding of how gyrase acts in the test tube, to how it behaves in the native environment inside living cells. While the super-resolution techniques used in this study cannot rival the atomic-level precision of structural biology studies, placing limitations on the extent of what we can really know about the activity of any individual gyrase enzyme, they offer an order of magnitude better spatial resolution than the standard optical resolution limit, and come with the substantive advantage that it is performed in living cells and thus allows us to answer questions which are impossible to answer with structural biology or in vitro biochemical techniques alone, such as 'how many gyrase act in proximity to the replication fork?'

Based on PALM and Slimfield analysis we estimate that an average of ~600 gyrase per cell are present of which ~300 are tightly DNA-bound and presumably performing catalysis. We find that gyrase forms foci which colocalize with replisomes and comprise on average of ~10 gyrase enzymes. In agreement with this, the fraction of DNA-bound gyrase is reduced by only a few % in cells that had either not yet initiated replication or had terminated replication but not divided. Despite the regions with the highest gyrase occupancy being close to the replisome, the vast majority of gyrase are immobile elsewhere on the chromosome. In a cell containing two replisomes there are at least ~1000 transcribing RNAPs, introducing (+) supercoils with an overall rate up to 30-fold higher than replication (~6000 compared to ~200 supercoils/s) (6,22). Since we find only ~20 out of 300 immobile gyrase are involved in relaxation of (+) supercoils introduced by replication, we expected the ~280 remaining to participate in relaxation of (+) supercoils introduced by transcription. We find that the fraction of immobile gyrase is reduced only modestly after transcription is blocked with rifampicin, indicating that the primary activity of gyrase is instead directed towards maintaining a steady-state level of (-) supercoiling, with a caveat that rifampicin itself has a major effect on nucleoid organization through decompaction (22), which may influence gyrase activities in an unknown way. Since the time taken to transcribe an average gene is short, it is inevitable that some of the (+) and (-) supercoiling created during transcription is cancelled out after RNAP dissociation. Similarly, on highly-expressed genes (+) supercoils produced ahead of multiple RNAPs will be neutralized by (-) supercoils introduced behind. Our results show that gyrase activity should not be considered as merely removing (+) supercoils to ensure unimpeded progression of transcription and replication, but contributes to multiple interdependent processes affecting global chromosome organization and segregation.

During replication of the chromosome over 220,000 catalytic events by the combined action of topo IV and gyrase must be performed, with gyrase removing (+) supercoils ahead of the replication fork and topo IV decatenating interlinked daughter chromosomes caused by diffusion of (+) supercoils behind the fork. These processes can occur simultaneously, yet the division of catalytic events between gyrase and topo IV during replication remains to be determined. Unlike gyrase, topo IV does not form foci in the proximity of the replisome (16,39,54). Nevertheless, blocking of topo IV prevents decatenation-segregation of all loci tested (11,16), demonstrating that the replisome can rotate and introduce precatenanes. Indeed, recent findings that most components of the replisome turnover every few seconds (55), suggest that the replisome is unlikely to be a barrier to replication fork rotation. The copy number of topo IV is much lower than gyrase; our previous measurements of topo IV under the same growth conditions as this study, showed that ~30 DNA-bound enzymes are present per cell, and the action of 1/3 of these are dependent on ongoing replication, indicating that during replication ~10 topo IVs are performing decatenation per cell (~5 per replication fork), most of which will be distal from the progressing forks since decatenation takes ~12 min (16).

The combined action of ~5 topo IV and ~10 gyrase enzymes per replication fork is nearly 10-fold lower that the number theoretically needed to keep up with replication rate, given the catalytic rate for both enzymes, which has been measured at ~1 supercoil/s. Importantly, topo IV is unlikely to decatenate processively, since *in vitro* topo IV acts distributively on (-) supercoils (with the same local topology as right-handed replicative catenanes) (56,57), confirmed by our previous measurements of topo IV dwell times (16). In contrast, gyrase can remove (+) supercoils processively *in vitro* (17-19), consistent with our observations that its dwell time significantly increases close to the replisome. Previous *in vitro* measurements of the processive catalytic rate were the same as for distributive catalysis (1 supercoil/s), and thus remains insufficient to account for the rate of supercoils introduced by each replisome (up to 100 supercoil/s). Intriguingly, a recent single-molecule *in vitro* study suggests that processive relaxation of (+) supercoils by *B.*

*anthracis* gyrase may be faster than previously measured for *E. coli* gyrase (18), with mean of ~6 supercoils/s (19), though with individual bursts of catalysis measured as high as 107±23 supercoils/s. Therefore, we suggest that the acute topological problem introduced by replication is primarily dealt with by gyrase enzymes performing processive catalysis to remove (+) supercoils ahead of replication, possibly at a higher rate than 1 supercoils/s, and we speculate that when gyrase fails to remove sufficient (+) supercoiling, replisome rotation is induced forming a substrate for topo IV behind the fork. However, it remains to be established whether *E. coli* gyrase *in vivo* can perform bursts of processive catalysis at higher rates than 1 supercoil/s.

The *E. coli* chromosome is organized into looped topological domains (8,21,58), within which supercoils can rapidly diffuse (5) and thus may delimit gyrase activity. Since the global net supercoiling of the chromosome is (-), most DNA loops will be relaxed or (-) supercoiled, and gyrase binding to these regions will perform a single round of catalysis. Our data suggest that local supercoiling may strongly influence gyrase off-rate, as we find with replication proximal gyrase remaining immobile for >8s. Since the fork progresses at a rate of up to 1000 bp/s this would require initially binding ~10 kbp ahead of the fork to avoid collisions rather than directly ahead of it. This predicts a displacement of gyrase foci in relation to replisome position. Indeed, Slimfield analysis (Figure 3C) showed that gyrase and replisome foci are displaced by ~100 nm. Therefore, diffusing (+) supercoils may promote processive catalysis of gyrase bound many kbp away from replication, which could help to protect against detrimental gyrase-fork collisions.

Together, our results show that *in vivo* a small number of gyrase acting processively ensures unimpeded progression of the replisome, while a majority of gyrase is involved in maintaining steady-state levels of chromosome supercoiling.

**Data availability.** Data included in full in the main text and supplementary files. Raw data available from the authors.

**Author contributions**: M.S., D.J.S., M.C.L., and P.Z., designed research. M.S., A.J.M.W., E.K., J.G., J.-E.L., V.A.L., S.J.M., L.A.M., P.Z., performed experiments and analyzed data. M.S., D.J.S., M.C.L., A.M., and P.Z., wrote the paper.

**FUNDING**

This work was supported by: the National Science Centre Poland [2015/19/P/NZ1/03859 to P.Z.] and FNP [First TEAM/2016-1/9 to P.Z.]; the Medical Research Council [MR/K01580X/1 to M.L.]; the Biotechnology and Biological Sciences Research Council [BB/N006453/1 to M.L., BB/R001235/1 to M.L. A.M., J.-E.L., BB/J004561/1, BB/P012523/1 to A.M.]; The Wellcome Trust through the Centre for Future Health at University of York [204829 to A.J.M.W.]; The Wellcome Trust [099204/Z/12Z to D.J.S.] and via a Sir Henry Wellcome Fellowship [204684/Z/16/Z to M.S.]; the Leverhulme Trust [RP2013-K-017 to A.M.]; and a Junior Research Fellowship at Trinity College Oxford [to M.S].

**Figure legends**

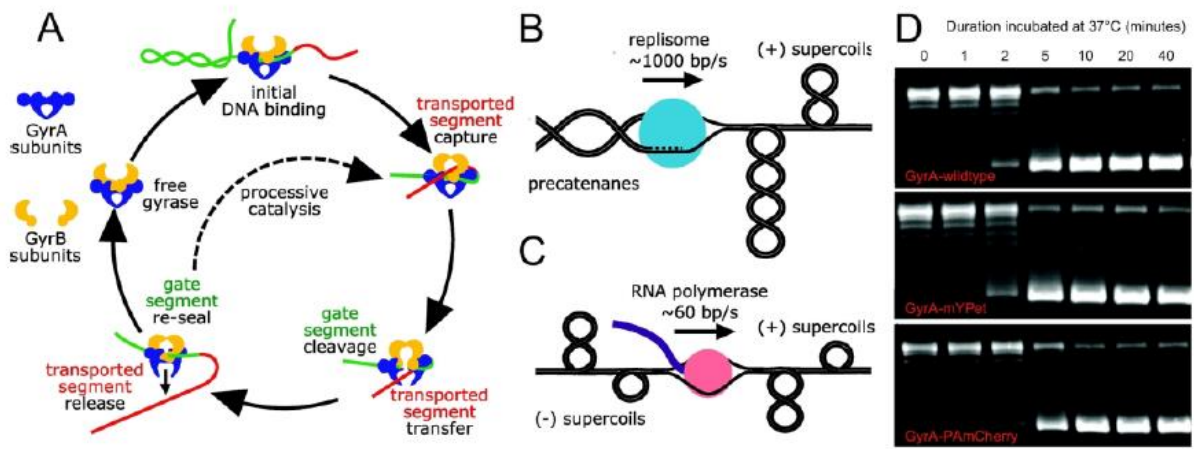

**Figure 1. The activity of gyrase**. A) DNA gyrase catalytic cycle. B) Replication introduces (+) supercoils ahead and precatenated DNA behind. Gyrase acts ahead of the fork while topo IV removes precatenanes behind. C) Gyrase removes (+) supercoiling from ahead of RNAP to ensure unperturbed transcription. D) Time course supercoiling assays presenting the activity of GyrA fusion proteins against the wild-type GyrA after different incubation periods at 37°C. Gyrase was incubated with relaxed pBR322 DNA in standard supercoiling assays. Samples were taken at the intervals indicated and loaded onto a 1% agarose gel and analysed by electrophoresis.

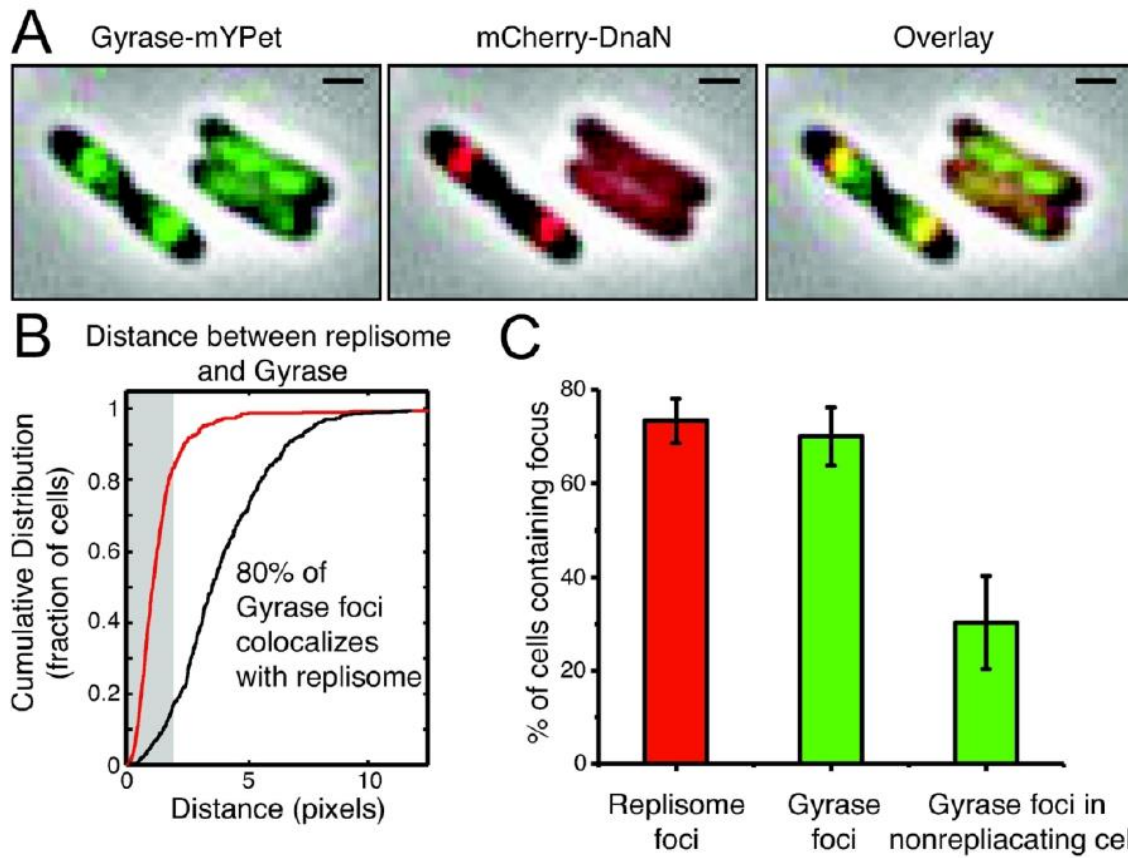

**Figure 2. Epifluorescence of *E. coli* gyrase.** A) Example cells with gyrase, fork marked with mCherry-DnaN, and overlay of signal from both channels; scale bar 1 µm. B) Cumulative distributions of distances between centroids of fork foci and brightest gyrase pixels in each cell (red), or a randomly simulated position (black). Colocalization (gray shaded rectangle) defined as when the fork centroid is ≤ 2 pixels (256 nm) from the brightest gyrase pixel. C) % of cells from population containing fork or gyrase foci plotted as a histogram. SD error bars from N=3 experiments.

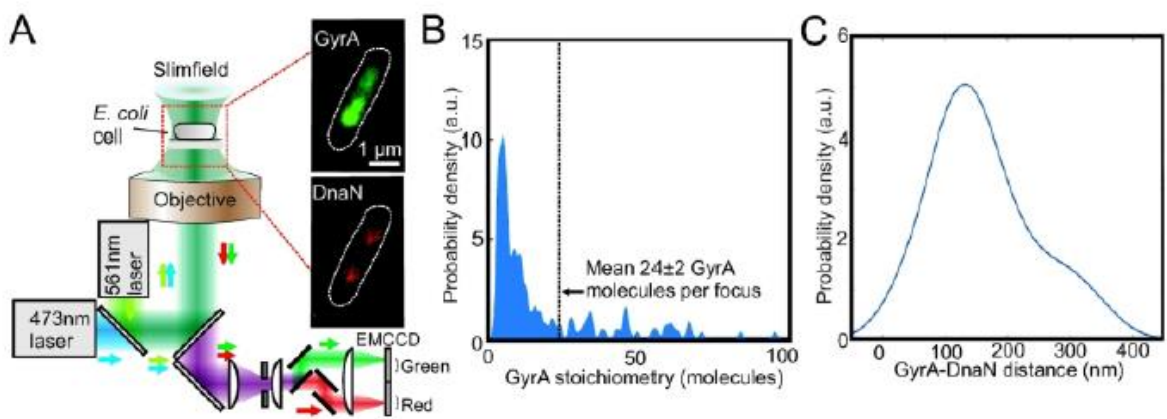

**Figure 3. GyrA form foci of a few tens of molecules.** A) Dual-color Slimfield

enables single-molecule tracking in two separate color channels with millisecond sampling, for the strain GyrA-mYPet:DnaN-mCherry, cell outline indicated (white dash). B) Stoichiometry distribution rendered as a kernel density estimate(38) for all detected GyrA-mYPet foci, mean (±SEM) indicated for all GyrA, kernel width 0.7 molecules. C) Distribution of displacements between foci centers for colocalized DnaN and GyrA rendered as a kernel density estimate, kernel width 40 nm. Data acquired from 72 foci using N=35 cells.

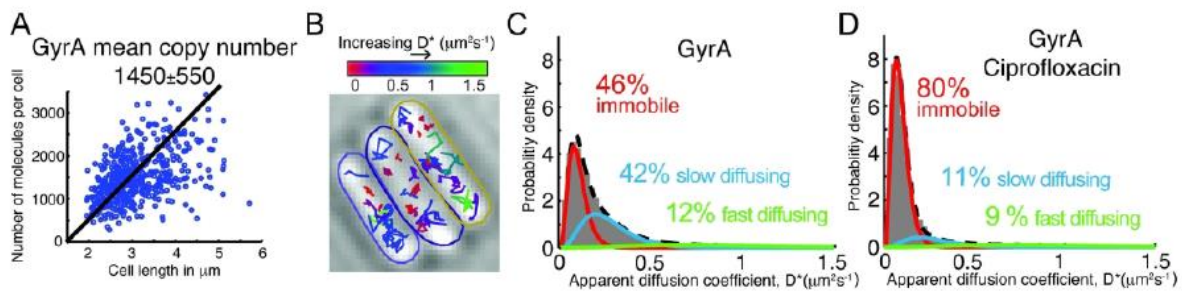

**Figure 4. Intracellular characterization of *E. coli* gyrase.** A) Copy number of GyrA in exponentially growing culture. B) Selected tracks colored according to apparent diffusion coefficient (*D**) of individual GyrA. C) Distribution of *D** for 85529 tracked GyrA. D) Distribution of *D** for 30813 GyrA treated for 10 min with 10 µg/ml ciprofloxacin.

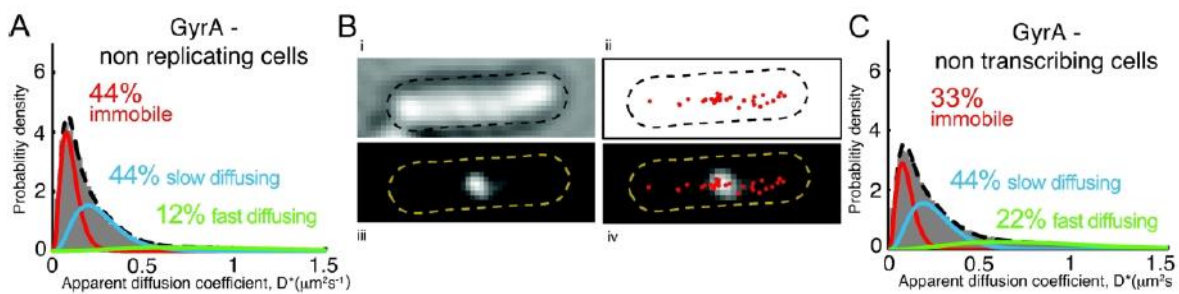

**Figure 5. Effect of replication and transcription on gyrase mobility.** A) Distribution of *D** for 16597 tracks in cells without fork foci. B) Cell (i - brightfield) with mean position of immobile molecules (ii) and position of fork marker mYPet-DnaN (iii). iv) Superimposed images of ii and iii. C) Distribution of *D** for 41632 GyrA in cells treated for 30 min with 50 µg/ml rifampicin.

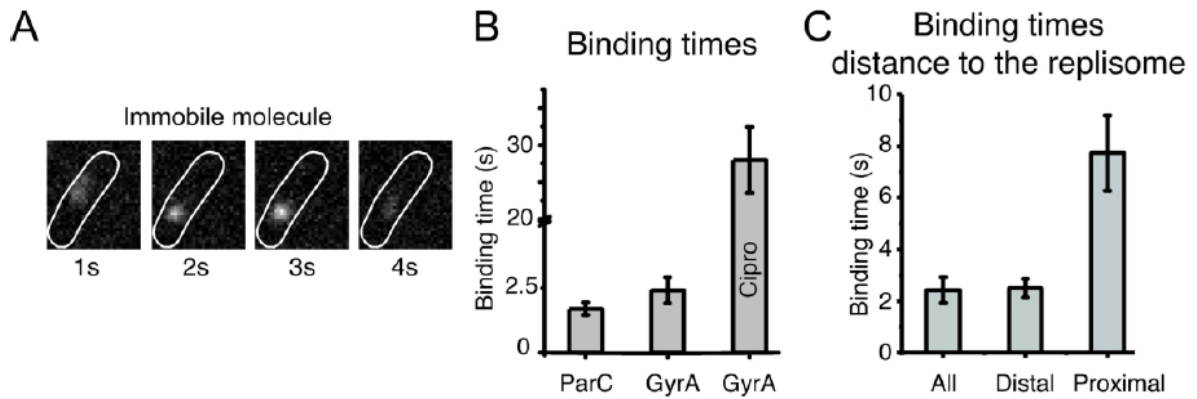

**Figure 6. Binding times of gyrase inside live cells**. A) PALM images of an example cell imaged with 1 s exposure times; only immobile GyrA-PAmCherry produce distinct foci, while mobile GyrA are blurred and produce signal below the detection threshold. B) Photobleaching-corrected binding times extracted from 1s exposures of GyrA-PAmCherry, topo IV subunit (ParC-PAmCherry) and GyrA after 10 min treatment with 10 µg/ml of ciprofloxacin. C) Photobleaching-corrected binding times for GyrA, dependent on the distance from fork, categorized as proximal (<200 nm), distal (≥200 nm) or all binding events.

**Movie 1. Legend** Example Slimfield GyrA-mYpet (yellow) fluorescence photobleaching. Time in ms shown, scale bar 1 µm.

# Supplementary Material & Methods and Supplementary Figures

**Increased activity of DNA gyrase near replication forks revealed by *in vivo* single-molecule imaging**

**Authors:**

Mathew Stracy[a,1], Adam J.M. Wollman[b,1], Elzbieta Kaja[e], Jacek Gapinski[c], Ji-Eun Lee[b], Victoria A Leek[d], Shannon J. McKie[d], Lesley A. Mitchenall[d], Anthony Maxwell[d], David J. Sherratt[a], Mark C. Leake[b,2], Pawel Zawadzki[a,c,2]

**Affiliations:**

[a]Department of Biochemistry, University of Oxford, South Parks Road, Oxford, OX1 3QU, United Kingdom.

[b]Biological Physical Sciences Institute (BPSI), Departments of Physics and Biology, University of York, York YO10 5DD, United Kingdom

[c]Molecular Biophysics Division, Faculty of Physics, A. Mickiewicz University,

Umultowska 85, 61-614 Poznan, Poland

[d]Department of Biological Chemistry, John Innes Centre, Norwich Research Park, Colney, Norwich NR4 7UH, United Kingdom

[e]NanoBioMedical Centre, Adam Mickiewicz University, Umultowska 85, 61-614 Poznan, Poland.

[1]These authors contributed equally

[2]Co-corresponding authors. Email: zawadzki@amu.edu.pl or mark.leake@york.ac.uk

**Bacterial strains**

All strains were derivatives of *Escherichia coli* K-12 AB1157 (1). The oligonucleotides used for replacement of genes with C-terminal mYPet fusions by λ-Red recombination (2) are shown in Table S1. PCRs were performed with the template plasmid pROD10, containing the sequence for the monmentic YPet fluorescent proetin preceeeded by a flexible 11 amino acid linker (SAGSAAGSGEF), and followed by an frt-flanked kanamycin resistance gene (*kan<sup>r</sup>*). For PAmCherry fusions the same oligo sets were used with the template plasmid pROD85 containing PAmCHerry instead of mYPet. For multiple insertions of modified genes, the *kan<sup>r</sup>* gene was removed using site-specific recombination through expression of the Flp recombinase from plasmid pCP20 (2). Correct insertion of the fragment into the chromosome was evaluated by PCR using primers flanking the insertion site.

| Oligonucleotides ||
|---|---|
| Name | Sequence |
| gyrApamcherryfor | GGACGATGAAATCGCTCCGGAAGTGGACGTTGACGACGAGCCAGAAGAAGAATCG GCT GGC TCC GCT GCT GGT TC |
| gyrApamcherryrev | TCAATTCAAACAAGGGAGATAGCTCCCTTTTGGCATGAAGAAGTAAAATTAGAGGATCCCATATGAATATCCTCC |
| gyrBpamcherryrev | GCCGTGCGTTTATTGAAGAGAACGCCCTGAAAGCGGCGAATATCGATATTTCG GCT GGC TCC GCT GCT GGT TC |

| gyrBpamcherryfor | GCCTGATAAGCGTAGCGCATCAGGCACGCTCGCATGGTTAGCGCCATTAGAGGATCCCATATGAATATCCTCC |

**Table S1.**

**Sample preparation**

Strains were streaked onto LB plates containing appropriate antibiotics. Single colonies were inoculated into M9 growth media with a glycerol carbon source (0.2%) and grown overnight at 37ºC to $A_{600}$ 0.4-0.6, then diluted into fresh M9 and grown to $A_{600}$ 0.1. Cells were centrifuged and immobilized on agarose pads between two glass coverslips. For PALM microscopy 0.17 mm thickness coverslips were cleaned of any background fluorescent particles before use by heating in an oven to 500ºC for 1 h. For Slimfield microscopy BK7 coverslip were plasma-cleaned before use. 1% agarose pads were prepared by mixing 2% low-fluorescence agarose (Bio-Rad) in $dH_2O$ 1:1 with 2x M9 growth medium. Where indicated cells were incubated with, 1 µg/ml ciprofloxacin for 10 minutes prior to imaging, or 50 µg/ml rifampicin for 30 minutes prior to imaging.

**Epifluorescence microscopy and colocalization analysis**

Wide-field epifluorescence microscopy was performed using an Eclipse TE2000-U microscope (Nikon), equipped with a 100x/NA1.4 oil PlanApo objective and a Cool-Snap $HQ^2$ CCD, and using Nikon NIS-Elements software for image acquisition. A chromosomally encoded mCherry-DnaN fusion protein was used as a marker for the replisome (3,4).

For colocalization analysis cell outlines were first delineated from a phase image using the MicrobeTracker software, creating a "mesh" for each cell, within which each pixel is characterized by a specific x,y coordinate. The positions of foci formed by mCherry-DnaN were established with Gaussian fitting described in the section titled ' Localization and tracking'. Since GyrA-mYPet did not form well-defined diffraction-limited foci, we determined the brightest pixel of GyrA-mYPet signal within each cell, as described in ref (6). It should be noted that the Gaussian localization analysis for mCherry-DnaN can identify multiple fluorescent foci within one cell or none at all, but the brightest pixel analysis finds exactly one pixel with the highest intensity for GyrA.

The pairwise distances between the brightest GyrA pixel and the nearest DnaN localization was calculated in Matlab (Mathworks) as described in ref (5). To determine the distribution of distances expected from an entirely random localization of GyrA, we also calculated distances between a pixel randomly positioned within the cell and the nearest DnaN focus. A threshold of 2 pixels (258 nm) was chosen to define colocalization.

**Photoactivated Localization Microscopy**

Live cell single-molecule-tracking PhotoActivated Localization Microscopy (PALM) was performed on a custom-built total internal reflection fluorescence (TIRF) microscope built around the Rapid Automated Modular Microscope (RAMM) System (ASI Imaging). Photoactivatable mCherry activation was controlled by a 405 nm laser and excitation with 561 nm. All lasers were provided by a multi-laser engine (iChrome MLE, Toptica). At the fibre output, the laser beams were collimated and focused (100x oil immersion objective, NA 1.4, Olympus) onto the sample under an angle allowing for highly inclined thin illumination (6). Fluorescence emission was filtered by a dichroic mirror and notch filter (ZT405/488/561rpc & ZET405/488/561NF, Chroma). PAmCherry emission was projected onto an EMCCD camera (iXon Ultra, 512x512 pixels, Andor). The pixel size was 96 nm. Transmission illumination was provided by an LED source and condenser (ASI Imaging). Sample position and focus were controlled with a motorized piezo stage, a z-motor objective mount, and autofocus system (MS-2000, PZ-2000FT, CRISP, ASI Imaging). PALM movies were aquired with a frame time of 15.48 ms. For colocalization analysis snapshots with 488 nm excitation were performed prior to PALM imaging.

**Localization and tracking**

PALM data for single-molecule-tracking analysis was localized using custom-written MATLAB software (MathWorks): fluorophore images were identified for localization by band-pass filtering and applying an intensity threshold to each frame of the super-resolution movie. Candidate positions were used as initial guesses in a two-dimensional elliptical Gaussian fit for high-precision localisation. Free fit parameters were x-position, y-position, x-width, y-width, elliptical rotation angle, intensity, background. Single-particle tracking analysis was performed by adapting the MATLAB implementation of the algorithm described in ref (7). Positions were linked to a track if they appeared in consecutive frames within a window of 5 pixels (0.48 µm). In rare cases when multiple localizations fell within the tracking radius, tracks were linked such that the sum of step distances was minimized. We used a 'memory' parameter of 1 frame to allow for transient (1 frame) disappearance of the fluorophore image within a track due to blinking or missed localisation.

**Molecule counting**

We counted the total number of GyrA or GyrB molecules by recording long movies (50000 frames), until no further activation was observed. Cells were segmented from transmission images using MicrobeTracker (8). Localizations within cell boundaries were tracked and the number of tracked molecules per cell established. We note that the copy numbers presented here may be underestimates of the true copy numbers, since only 49% of PAmCherry were shown to be photoactivatable in studies in eukaryotic cells (9).

**Measuring the diffusion of PAmCherry labeled proteins**.

We determined the mobility of each molecule by calculating an apparent (or nominal) diffusion coefficient, *D\**, from the one-step mean-squared displacement (MSD) of the track using:

$$D^* = \frac{1}{4n\Delta t} \sum_{i=1}^{n} [x(i\Delta t) - x(i\Delta t + \Delta t)]^2 + [y(i\Delta t) - y(i\Delta t + \Delta t)]^2$$

Where $x(t)$ and $y(t)$ are the coordinates of the molecule at time $t$, the frame time of the camera is $\Delta t$, and $n$ is the number of steps in the trajectory. Tracks shorter than $n = 4$ steps long were discarded for this analysis because the higher uncertainty in *D\** value.

For a molecule with apparent diffusion coefficient *D*, the probability distribution of obtaining a single-molecule *D\** value, $x$, is given by (10) :

$$f(x; D, n) = \frac{(n/D)^n x^{n-1} e^{-nx/D}}{(n-1)!}$$

Where $n$ is the number of steps in the trajectory. In order to determine the apparent diffusion coefficient, *D*, from the population of individual single-molecule *D\** values, longer tracks were truncated after 5[th] localization (i.e. $n = 4$). The *D\** distribution, $x$, was then fitted to the $n = 4$ analytical expressionequation:

$$f(x; D) = \frac{(4/D)^4 x^3 e^{-4x/D}}{6}$$

Fits were performed using maximum likelihood estimation in MATLAB, and errors were estimated as the SD in each estimated parameter using bootstrap resampling with 100 resamples, rounded up to the nearest 0.01 µm$^2$s$^{-1}$. A single species model fits poorly to the data (Supplementary Fig. 1c). We reasoned that at least two species with different mobilities are present: mobile molecules diffusing and binding only transiently to DNA, and immboile molecules bound to DNA for the entire trajectory. We therefore introduced a second species:

$$f(x; D_1, D_2, A) = \frac{A(4/D_1)^4 x^3 e^{-4x/D_1}}{6} + \frac{(1-A)(4/D_2)^4 x^3 e^{-4x/D_2}}{6}$$

Where $D_1$ and $D_2$ are the diffusion coefficients of the two different species, and $A$ and $1 - A$ are the fraction of molecules found in each state.

The localisation uncertainty in each measurement, $\sigma_{loc}$, manifests itself as a positive offset in the *D\** value of $\sigma_{loc}^2/\Delta t$(11). Based on the estimated localisation uncertainty of ~40 nm for our measurements, we expected a positive shift in the mean *D\** value of immobile molecules to ~0.1 µm$^2$s$^{-1}$.

**Estimating colocalization with the replisome**

The replisome position was esablished using a mYPet-DnaN fusion. Snapshots of mYPet-DnaN were taken prior to PALM imaging, and the exact position estimated using guassian fitting as described in the section 'Localization and tracking'. Cells were segmented based on transmission images using MicrobeTracker, and the number of PALM localization within each cell outline was determined. The pairwise distances between centroid positions of DnaN and all GyrA PALM localizations within each cell was determined using the pdist2 function in Matlab, and the fraction located within 200 nm was determined. The mean colocalized fraction was determined from all cells from the data set (containing at least 100 cells), and the SEM established from the means of five experimental data sets.

**Measuring long-lasting binding events**

PALM movies to measure long duration binding events were recorded at low continuous 561 nm excitation intensities using 1 s exposure times(12,13). At this exposure times mobile GyrA-PAmCherry molecules are motion blurred over a large fraction of the cell, whereas immobile GyrA-PAmCherry molecules still appear as point sources, producing a diffraction limited spot. Elliptical Gaussian fitting was used as described in the 'Localization and tracking' section. Bound and mobile molecules were distinguished by the width of the elliptical fits, with thresholds short axis-width < 160 nm and long axis-width < 200 nm to identify bound molecules. The probability of observing a particular on-time is the product of the underlying binding-time probability and the bleaching probability. The bleaching-time distributions were measured independently using MukB-PAmCherry which has a binding time >> bleaching time. On-time and bleaching-time distributions were fitted with single-exponential functions to extract exponential-time constants $t_{on}$ and $t_{bleach}$, and the binding-time constant was calculated by $t_{bound} = t_{on} \cdot t_{bleach} / (t_{bleach} - t_{on})$. Stochastic photoactivation of GyrA-PAmCherry molecules before or during binding events does not influence our measurement, because the observed binding times follow an exponential distribution and are therefore memoryless. The MukB-PAmCherry bleaching time constant, $t_{bleach} = 1.16 \pm 0.04$. The uncorrected $t_{on}$ time constants from 7 experimental repeats are shown in Table S2.

To determine binding times near the replisome, snapshots of mYPet-DnaN were taken prior to PALM imaging. DnaN foci were localized with Gaussian fitting and GyrA trajectories within 200 nm of a foci were used for binding time analysis. As a control, the binding times within 200 nm of mid-cell (where the replisome is expected to assemble/diasassemble) were determined in cells lacking DnaN foci. Mid-cell position was determined from segmenting the transmission image.

| Uncorrected on-time measurements in seconds |
|---|

| All tracks | Tracks >200nm from the replisome | Tracks <200nm from the replisome |
|---|---|---|
| 0.781 | 0.740 | 1.087 |
| 0.790 | 0.758 | 1.168 |
| 0.879 | 0.861 | 1.004 |
| 0.823 | 0.802 | 1.046 |
| 0.819 | 0.791 | 1.029 |
| 0.894 | 0.871 | 1.113 |
| 0.909 | 0.887 | 1.132 |

**In vitro DNA supercoiling assay.**

Wildtype and fluorescently tagged GyrA and GyrB subunits were purified according to standard protocols (14). Supercoiling assays were carried out as before (14). Briefly, a 1.5 µL aliquot of the 0.1 µM respective GyrA sample was added to 17 µL of $H_2O$, 4 µL of dilution buffer, 6 µL of assay buffer, 0.5 µL of relaxed DNA and 1 µL of GyrB (0.75 µM). This resulted in a final concentration of GyrA and GyrB of 5 nM and 25 nM, respectively. Full supercoiling activity was observed after 5-10 minutes for the GyrA wildtype, which remained consistent across the repeats. However, the activities of the two fusions were minimally lower but still comparable to the wild type.

**Slimfield image analysis**

Foci from Slimfield images were automatically detected and tracked using custom-written Matlab software discussed previously (15). In brief, bright foci were identified by image transformation and thresholding. The centroid of candidate foci were determined using iterative Gaussian masking and accepted if their intensity was greater than a signal to noise (SNR) of 0.4. Intensity was defined as the summed pixel intensity inside a 5 pixel circular region of interest (ROI) corrected for the background in an outer square ROI of 17x17 pixels. SNR was defined as the mean BG corrected pixel intensity in the circular ROI divided by the standard deviation in the square ROI. Foci were linked together into trajectories between frames if they were within 5 pixels of each other.

Stoichiometry was determined by fitting the first 3 intensity values of a foci to a straight line, using the intercept as the initial intensity and dividing this by the characteristic intensity of mYPet or mCherry. This characteristic intensity was determined from the distribution of foci intensity values towards the end of the photobleach confirmed by overtracking foci beyond their bleaching to generate individual photobleach steps of the characteristic intensity (Fig S2). Red and green images were aligned based on the peak of the 2D cross correlation between brightfield images using individual green channel image frame cross correlated against 10 frame average images from the red channel. Colocalisation between green and red foci and the probability of random colocalisation was determined as described previously(16).

Copy numbers were determined using the first excited mYPet image frame. The image was segmented and background corrected using the mean intensity from images of the wild type *E. coli* without mYPet but imaged using identical conditions. A model 'sausage function' *E. coli* shape was fitted to the segmented area using the minor and major radii. A model 3D point spread function was integrated over this volume and the molecular concentration determined by solving a set of linear equations for each pixel in the real, background corrected image and model convolved image (17).

**Supplementary References**

**Supplementary Figures**

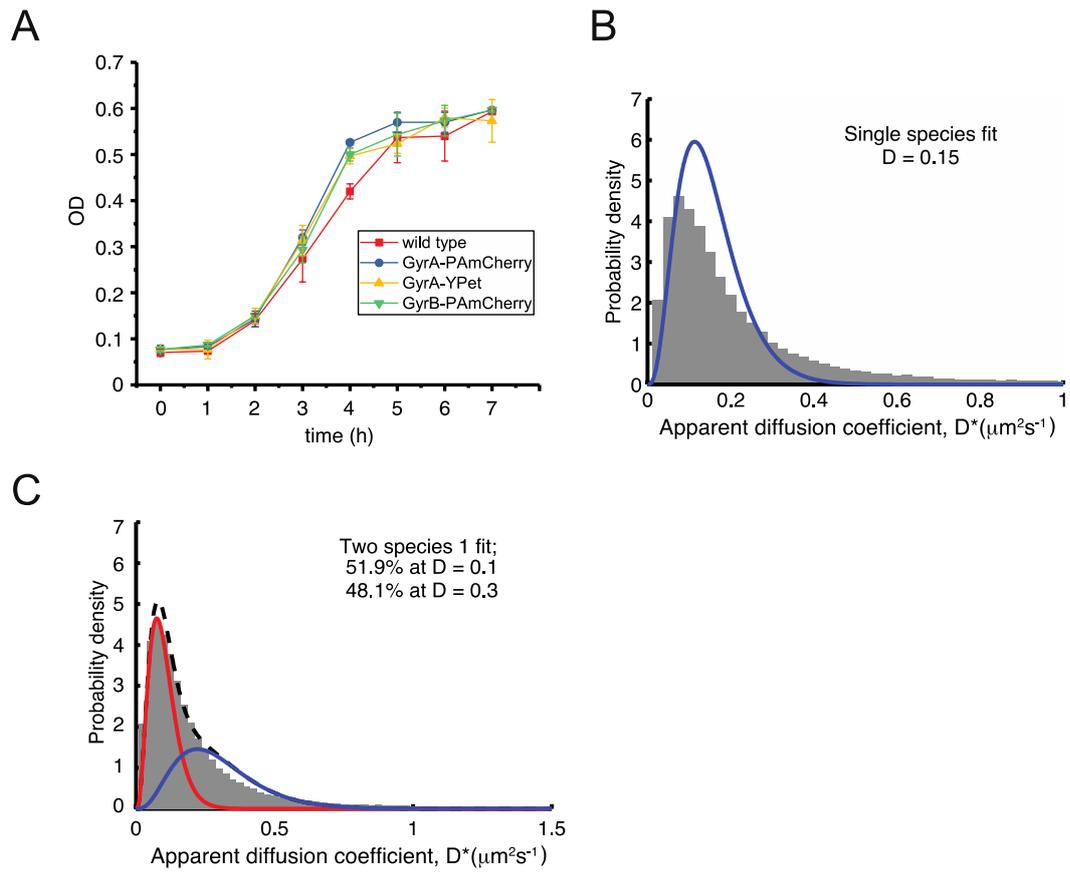

**Figure S1.** A) Growth curves of indicated strains in LB media at 37°C. B) Single-species fit to GyrA data. C) Double-species fit to GyrA data.

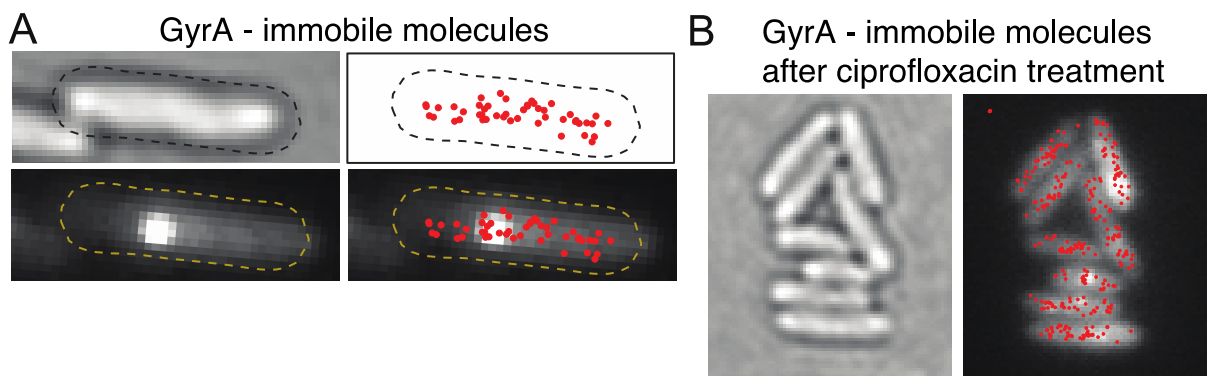

**Figure S2.** A) An example of the cell where no clear enrichment of GyrA close to the replisome was observed. Red dots represent mean position of immobile molecule. B) Group of cells after ciprofloxacin treatment (molecules treated for 10 min with 10 µg/ml ciprofloxacin) demonstrating that catalytically active gyrase are distributed throughout the chromosome.

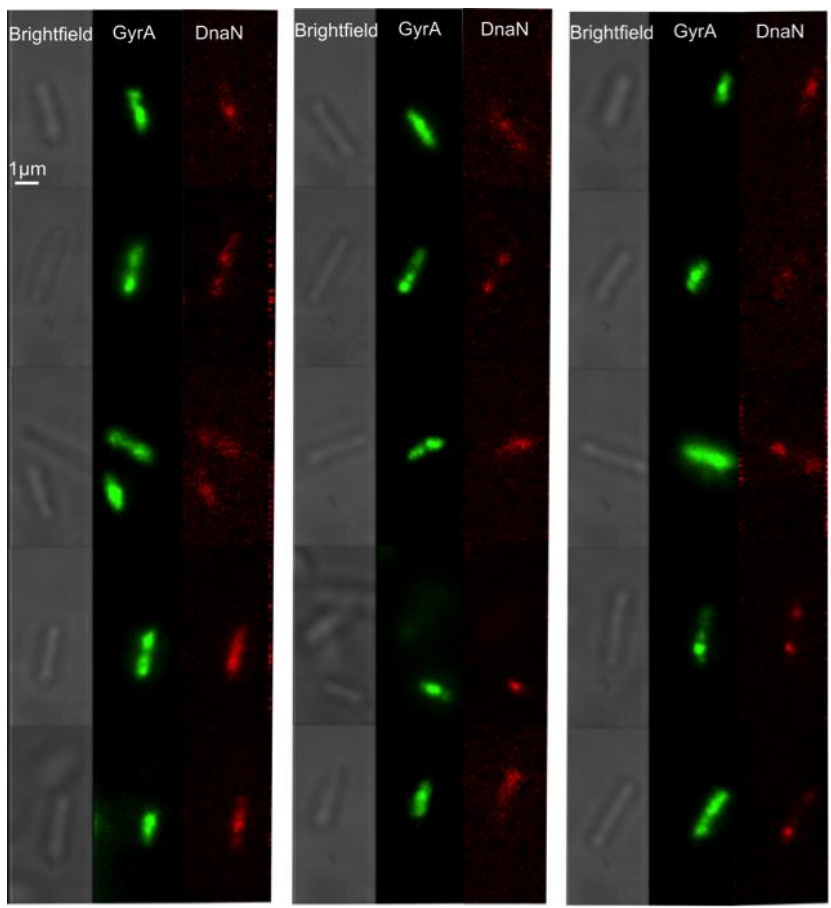

**Figure S3.** Brightfield and equivalent GyrA-mYPet and DnaN-mCherry dual-color Slimfield images (frame averages, from first five frames).

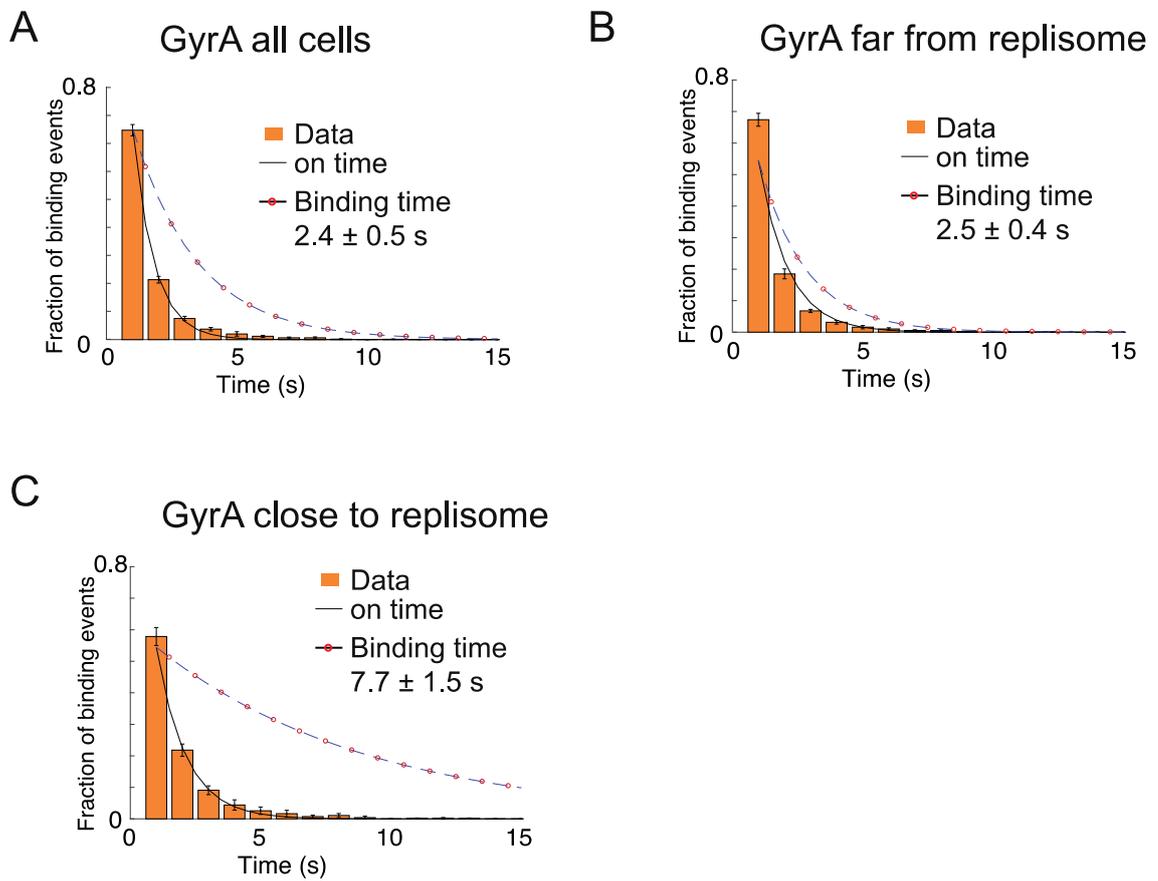

**Figure S4. Dwell times of GyrA.** On-time distributions for immobile GyrA-PAmCherry imaged with 1 s exposure times. Single exponential fits (solid lines) and photobleaching-corrected on time distributions (dashed circled lines). Photobleaching times were estimated by imaging, under the same conditions, cells with MukB-PAmCherry fusion, which has been shown to have a dwell time of ~50 s. Error bars shows S.E.M. of three experimental repeats.

# Supplementary Material & Methods and Supplementary Figures

# Single-molecule imaging of DNA gyrase activity in living *Escherichia coli*

**Authors:**

Mathew Stracy[a,1], Adam J.M. Wollman[b,1], Elzbieta Kaja[e], Jacek Gapinski[c], Ji-Eun Lee[b], Victoria A Leek[d], Shannon J. McKie[d], Lesley A. Mitchenall[d], Anthony Maxwell[d], David J. Sherratt[a], Mark C. Leake[b,2], Pawel Zawadzki[a,c,2]

**Affiliations:**

[a]Department of Biochemistry, University of Oxford, South Parks Road, Oxford, OX1 3QU, United Kingdom.

[b]Biological Physical Sciences Institute (BPSI), Departments of Physics and Biology, University of York, York YO10 5DD, United Kingdom

[c]Molecular Biophysics Division, Faculty of Physics, A. Mickiewicz University,

Umultowska 85, 61-614 Poznan, Poland

[d]Department of Biological Chemistry, John Innes Centre, Norwich Research Park, Colney, Norwich NR4 7UH, United Kingdom

[e]NanoBioMedical Centre, Adam Mickiewicz University, Umultowska 85, 61-614 Poznan, Poland.

[1]These authors contributed equally

[2]Co-corresponding authors. Email: zawadzki@amu.edu.pl or mark.leake@york.ac.uk

**Bacterial strains**

All strains were derivatives of *Escherichia coli* K-12 AB1157 (1). The oligonucleotides used for replacement of genes with C-terminal mYPet fusions by λ-Red recombination (2) are shown in Table S1. PCRs were performed with the template plasmid pROD10, containing the sequence for the monmentic YPet fluorescent proetin preceeeded by a flexible 11 amino acid linker (SAGSAAGSGEF), and followed by an frt-flanked kanamycin resistance gene (*kan*[r]). For PAmCherry fusions the same oligo sets were used with the template plasmid pROD85 containing PAmCHerry instead of mYPet. For multiple insertions of modified genes, the *kan*[r] gene was removed using site-specific recombination through expression of the Flp recombinase from plasmid pCP20 (2). Correct insertion of the fragment into the chromosome was evaluated by PCR using primers flanking the insertion site.

| Oligonucleotides | |
|---|---|
| Name | Sequence |
| gyrApamcherryfor | GGACGATGAAATCGCTCCGGAAGTGGACGTTGACGACGAGCCAGAAGAAGAATCG GCT GGC TCC GCT GCT GGT TC |
| gyrApamcherryrev | TCAATTCAAACAAGGGAGATAGCTCCCTTTTGGCATGAAGAAGTAAAATTAGAGGATCCCATATGAATATCCTCC |
| gyrBpamcherryrev | GCCGTGCGTTTATTGAAGAGAACGCCCTGAAAGCGGCGAATATCGATATTTCG GCT GGC TCC GCT GCT GGT TC |
| gyrBpamcherryfor | GCCTGATAAGCGTAGCGCATCAGGCACGCTCGCATGGTTAGCGCCATTAGAGGATCCCATATGAATATCCTCC |

**Table S1.**

**Sample preparation**

Strains were streaked onto LB plates containing appropriate antibiotics. Single colonies were inoculated into M9 growth media with a glycerol carbon source (0.2%) and grown overnight at 37ºC to $A_{600}$ 0.4-0.6, then diluted into fresh M9 and grown to $A_{600}$ 0.1. Cells were centrifuged and immobilized on agarose pads between two glass coverslips. For PALM microscopy 0.17 mm thickness coverslips were cleaned of any background fluorescent particles before use by heating in an oven to 500ºC for 1 h. For Slimfield microscopy BK7 coverslip were plasma-cleaned before use. 1% agarose pads were prepared by mixing 2% low-fluorescence agarose (Bio-Rad) in $dH_2O$ 1:1 with 2x M9 growth medium. Where indicated cells were incubated with, 1 µg/ml ciprofloxacin for 10 minutes prior to imaging, or 50 µg/ml rifampicin for 30 minutes prior to imaging.

**Epifluorescence microscopy and colocalization analysis**

Wide-field epifluorescence microscopy was performed using an Eclipse TE2000-U microscope (Nikon), equipped with a 100x/NA1.4 oil PlanApo objective and a Cool-Snap $HQ^2$ CCD, and using Nikon NIS-Elements software for image acquisition. A chromosomally encoded mCherry-DnaN fusion protein was used as a marker for the replisome (3,4).

For colocalization analysis cell outlines were first delineated from a phase image using the MicrobeTracker software, creating a "mesh" for each cell, within which each pixel is characterized by a specific x,y coordinate. The positions of foci formed by mCherry-DnaN were established with Gaussian fitting described in the section titled ' Localization and tracking'. Since GyrA-mYPet did not form well-defined diffraction-limited foci, we determined the brightest pixel of GyrA-mYPet signal within each cell, as described in ref (6). It should be noted that the Gaussian localization analysis for mCherry-DnaN can identify multiple fluorescent foci within one cell or none at all, but the brightest pixel analysis finds exactly one pixel with the highest intensity for GyrA.

The pairwise distances between the brightest GyrA pixel and the nearest DnaN localization was calculated in Matlab (Mathworks) as described in ref (5). To determine the distribution of distances expected from an entirely random localization of GyrA, we also calculated distances between a pixel randomly positioned within the cell and the nearest DnaN focus. A threshold of 2 pixels (258 nm) was chosen to define colocalization.

**Photoactivated Localization Microscopy**

Live cell single-molecule-tracking PhotoActivated Localization Microscopy (PALM) was performed on a custom-built total internal reflection fluorescence (TIRF) microscope built around the Rapid Automated Modular Microscope (RAMM) System (ASI Imaging). Photoactivatable mCherry activation was controlled by a 405 nm laser and excitation with 561 nm. All lasers were provided by a multi-laser engine (iChrome MLE, Toptica). At the fibre output, the laser beams were collimated and focused (100x oil immersion objective, NA 1.4, Olympus) onto the sample under an angle allowing for highly inclined thin illumination (6). Fluorescence emission was filtered by a dichroic mirror and notch filter (ZT405/488/561rpc & ZET405/488/561NF, Chroma). PAmCherry emission was projected onto an EMCCD camera (iXon Ultra, 512x512 pixels, Andor). The pixel size was 96 nm. Transmission illumination was provided by an LED source and condenser (ASI Imaging). Sample position and focus were controlled with a motorized piezo stage, a z-motor objective mount, and autofocus system (MS-2000, PZ-2000FT, CRISP, ASI Imaging). PALM movies were aquired with a frame time of 15.48 ms. For colocalization analysis snapshots with 488 nm excitation were performed prior to PALM imaging.

**Localization and tracking**

PALM data for single-molecule-tracking analysis was localized using custom-written MATLAB software (MathWorks): fluorophore images were identified for localization by band-pass filtering and applying an intensity threshold to each frame of the super-resolution movie. Candidate positions were used as initial guesses in a two-dimensional elliptical Gaussian fit for high-precision localisation. Free fit parameters were x-position, y-position, x-width, y-width, elliptical rotation angle, intensity, background. Single-particle tracking analysis was performed by adapting the MATLAB implementation of the algorithm described in ref (7). Positions were linked to a track if they appeared in consecutive frames within a window of 5 pixels (0.48 µm). In rare cases when multiple localizations fell within the tracking radius, tracks were linked such that the sum of step distances was minimized. We used a 'memory' parameter of 1 frame to allow for transient (1 frame) disappearance of the fluorophore image within a track due to blinking or missed localisation.

**Molecule counting**

We counted the total number of GyrA or GyrB molecules by recording long movies (50000 frames), until no further activation was observed. Cells were segmented from transmission images using MicrobeTracker (8). Localizations within cell boundaries were tracked and the number of tracked molecules per cell established. We note that the copy numbers presented here may be underestimates of the true copy numbers, since only 49% of PAmCherry were shown to be photoactivatable in studies in eukaryotic cells (9).

**Measuring the diffusion of PAmCherry labeled proteins**.

We determined the mobility of each molecule by calculating an apparent (or nominal) diffusion coefficient, *D\**, from the one-step mean-squared displacement (MSD) of the track using:

$$D^* = \frac{1}{4n\Delta t} \sum_{i=1}^{n} [x(i\Delta t) - x(i\Delta t + \Delta t)]^2 + [y(i\Delta t) - y(i\Delta t + \Delta t)]^2$$

Where $x(t)$ and $y(t)$ are the coordinates of the molecule at time $t$, the frame time of the camera is $\Delta t$, and $n$ is the number of steps in the trajectory. Tracks shorter than $n = 4$ steps long were discarded for this analysis because the higher uncertainty in *D\** value.

For a molecule with apparent diffusion coefficient *D*, the probability distribution of obtaining a single-molecule *D\** value, $x$, is given by (10) :

$$f(x; D, n) = \frac{(n/D)^n x^{n-1} e^{-nx/D}}{(n-1)!}$$

Where $n$ is the number of steps in the trajectory. In order to determine the apparent diffusion coefficient, *D*, from the population of individual single-molecule *D\** values, longer tracks were truncated after 5th localization (i.e. $n = 4$). The *D\** distribution, $x$, was then fitted to the $n = 4$ analytical expressionequation:

$$f(x; D) = \frac{(4/D)^4 x^3 e^{-4x/D}}{6}$$

Fits were performed using maximum likelihood estimation in MATLAB, and errors were estimated as the SD in each estimated parameter using bootstrap resampling with 100 resamples, rounded up to the nearest 0.01 µm²s⁻¹. A single species model fits poorly to the data (Supplementary Fig. 1c). We reasoned that at least two species with different mobilities are present: mobile molecules diffusing and binding only transiently to DNA, and immboile molecules bound to DNA for the entire trajectory. We therefore introduced a second species:

$$f(x; D_1, D_2, A) = \frac{A(4/D_1)^4 x^3 e^{-4x/D_1}}{6} + \frac{(1-A)(4/D_2)^4 x^3 e^{-4x/D_2}}{6}$$

Where $D_1$ and $D_2$ are the diffusion coefficients of the two different species, and $A$ and $1-A$ are the fraction of molecules found in each state.

The localisation uncertainty in each measurement, $\sigma_{loc}$, manifests itself as a positive offset in the *D\** value of $\sigma_{loc}^2/\Delta t$(11). Based on the estimated localisation uncertainty of ~40 nm for our measurements, we expected a positive shift in the mean *D\** value of immobile molecules to ~0.1 µm²s⁻¹.

**Estimating colocalization with the replisome**

The replisome position was esablished using a mYPet-DnaN fusion. Snapshots of mYPet-DnaN were taken prior to PALM imaging, and the exact position estimated using guassian fitting as described in the section 'Localization and tracking'. Cells were segmented based on transmission images using MicrobeTracker, and the number of PALM localization within each cell outline was determined. The pairwise distances between centroid positions of DnaN and all GyrA PALM localizations within each cell was determined using the pdist2 function in Matlab, and the fraction located within 200 nm was determined. The mean colocalized fraction was determined from all cells from the data set (containing at least 100 cells), and the SEM established from the means of five experimental data sets.

**Measuring long-lasting binding events**

PALM movies to measure long duration binding events were recorded at low continuous 561 nm excitation intensities using 1 s exposure times(12,13). At this exposure times mobile GyrA-PAmCherry molecules are motion blurred over a large fraction of the cell, whereas immobile GyrA-PAmCherry molecules still appear as point sources, producing a diffraction limited spot. Elliptical Gaussian fitting was used as described in the 'Localization and tracking' section. Bound and mobile molecules were distinguished by the width of the elliptical fits, with thresholds short axis-width < 160 nm and long axis-width < 200 nm to identify bound molecules. The probability of observing a particular on-time is the product of the underlying binding-time probability and the bleaching probability. The bleaching-time distributions were measured independently using MukB-PAmCherry which has a binding time >> bleaching time. On-time and bleaching-time distributions were fitted with single-exponential functions to extract exponential-time constants $t_{on}$ and $t_{bleach}$, and the binding-time constant was calculated by $t_{bound} = t_{on} \cdot t_{bleach} / (t_{bleach} - t_{on})$. Stochastic photoactivation of GyrA-PAmCherry molecules before or during binding events does not influence our measurement, because the observed binding times follow an exponential distribution and are therefore memoryless. The MukB-PAmCherry bleaching time constant, $t_{bleach} = 1.16 \pm 0.04$. The uncorrected $t_{on}$ time constants from 7 experimental repeats are shown in Table S2.

To determine binding times near the replisome, snapshots of mYPet-DnaN were taken prior to PALM imaging. DnaN foci were localized with Gaussian fitting and GyrA trajectories within 200 nm of a foci were used for binding time analysis. As a control, the binding times within 200 nm of mid-cell (where the replisome is expected to assemble/diasassemble) were determined in cells lacking DnaN foci. Mid-cell position was determined from segmenting the transmission image.

| Uncorrected on-time measurements in seconds |
|---|

| All tracks | Tracks >200nm from the replisome | Tracks <200nm from the replisome |
|---|---|---|
| 0.781 | 0.740 | 1.087 |
| 0.790 | 0.758 | 1.168 |
| 0.879 | 0.861 | 1.004 |
| 0.823 | 0.802 | 1.046 |
| 0.819 | 0.791 | 1.029 |
| 0.894 | 0.871 | 1.113 |
| 0.909 | 0.887 | 1.132 |

**In vitro DNA supercoiling assay.**

Wildtype and fluorescently tagged GyrA and GyrB subunits were purified according to standard protocols (14). Supercoiling assays were carried out as before (14). Briefly, a 1.5 µL aliquot of the 0.1 µM respective GyrA sample was added to 17 µL of $H_2O$, 4 µL of dilution buffer, 6 µL of assay buffer, 0.5 µL of relaxed DNA and 1 µL of GyrB (0.75 µM). This resulted in a final concentration of GyrA and GyrB of 5 nM and 25 nM, respectively. Full supercoiling activity was observed after 5-10 minutes for the GyrA wildtype, which remained consistent across the repeats. However, the activities of the two fusions were minimally lower but still comparable to the wild type.

**Slimfield image analysis**

Foci from Slimfield images were automatically detected and tracked using custom-written Matlab software discussed previously (15). In brief, bright foci were identified by image transformation and thresholding. The centroid of candidate foci were determined using iterative Gaussian masking and accepted if their intensity was greater than a signal to noise (SNR) of 0.4. Intensity was defined as the summed pixel intensity inside a 5 pixel circular region of interest (ROI) corrected for the background in an outer square ROI of 17x17 pixels. SNR was defined as the mean BG corrected pixel intensity in the circular ROI divided by the standard deviation in the square ROI. Foci were linked together into trajectories between frames if they were within 5 pixels of each other.

Stoichiometry was determined by fitting the first 3 intensity values of a foci to a straight line, using the intercept as the initial intensity and dividing this by the characteristic intensity of mYPet or mCherry. This characteristic intensity was determined from the distribution of foci intensity values towards the end of the photobleach confirmed by overtracking foci beyond their bleaching to generate individual photobleach steps of the characteristic intensity (Fig S2). Red and green images were aligned based on the peak of the 2D cross correlation between brightfield images using individual green channel image frame cross correlated against 10 frame average images from the red channel. Colocalisation between green and red foci and the probability of random colocalisation was determined as described previously(16).

Copy numbers were determined using the first excited mYPet image frame. The image was segmented and background corrected using the mean intensity from images of the wild type *E. coli* without mYPet but imaged using identical conditions. A model 'sausage function' *E. coli* shape was fitted to the segmented area using the minor and major radii. A model 3D point spread function was integrated over this volume and the molecular concentration determined by solving a set of linear equations for each pixel in the real, background corrected image and model convolved image (17).

**Supplementary References**

**Supplementary Figures**

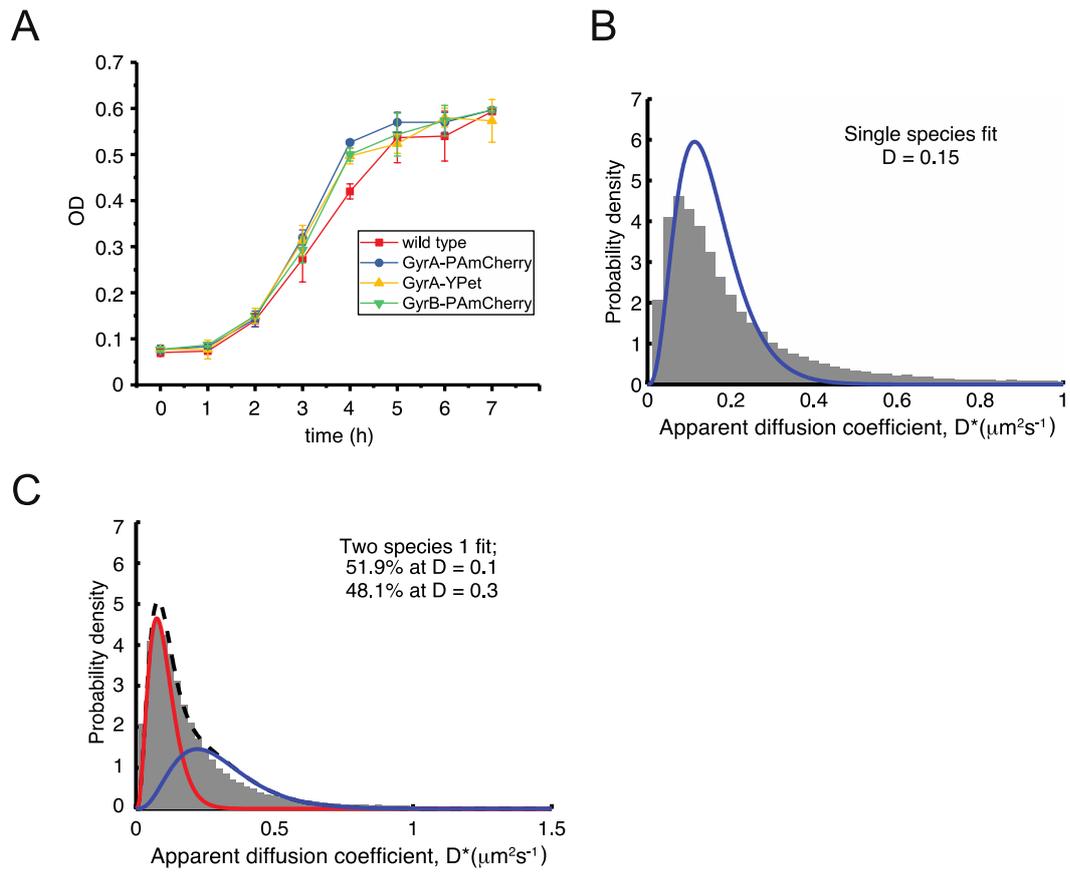

**Figure S1.** A) Growth curves of indicated strains in LB media at 37°C. B) Single-species fit to GyrA data. C) Double-species fit to GyrA data.

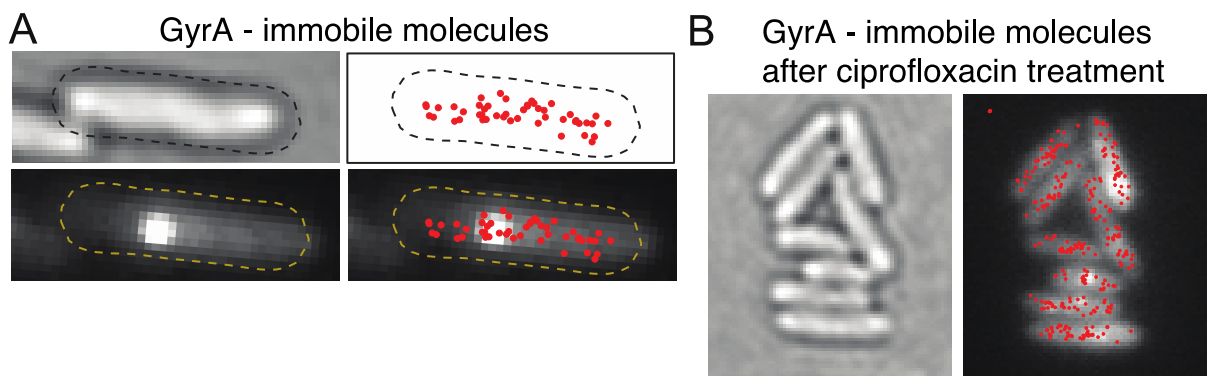

**Figure S2.** A) An example of the cell where no clear enrichment of GyrA close to the replisome was observed. Red dots represent mean position of immobile molecule. B) Group of cells after ciprofloxacin treatment (molecules treated for 10 min with 10 μg/ml ciprofloxacin) demonstrating that catalytically active gyrase are distributed throughout the chromosome.

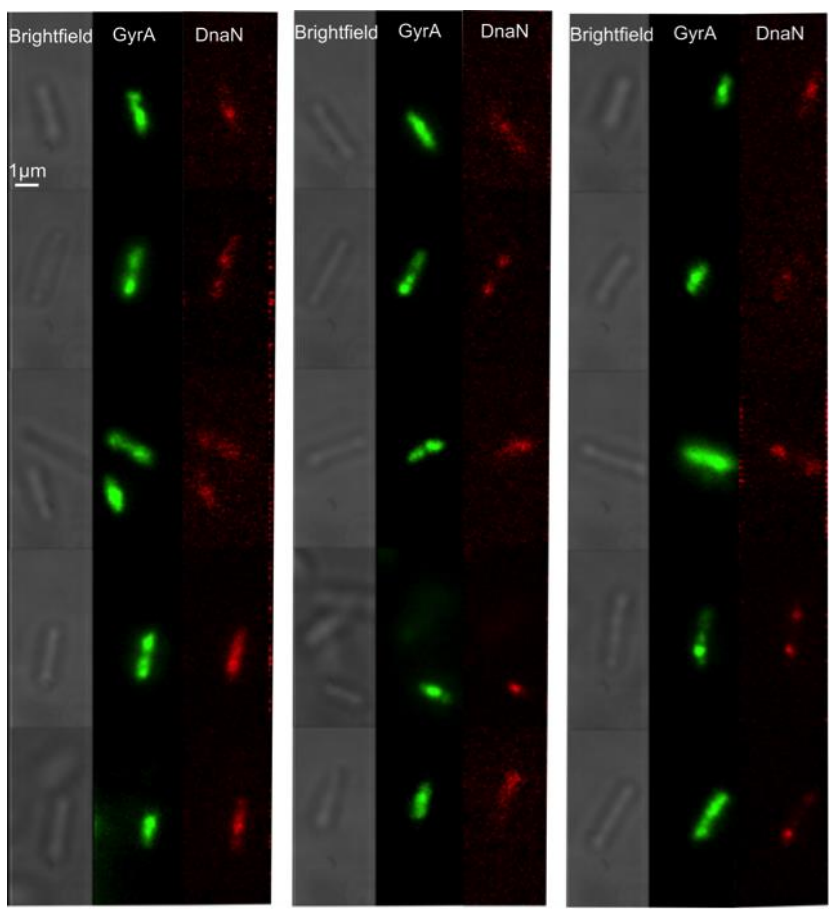

**Figure S3.** Brightfield and equivalent GyrA-mYPet and DnaN-mCherry dual-color Slimfield images (frame averages, from first five frames).

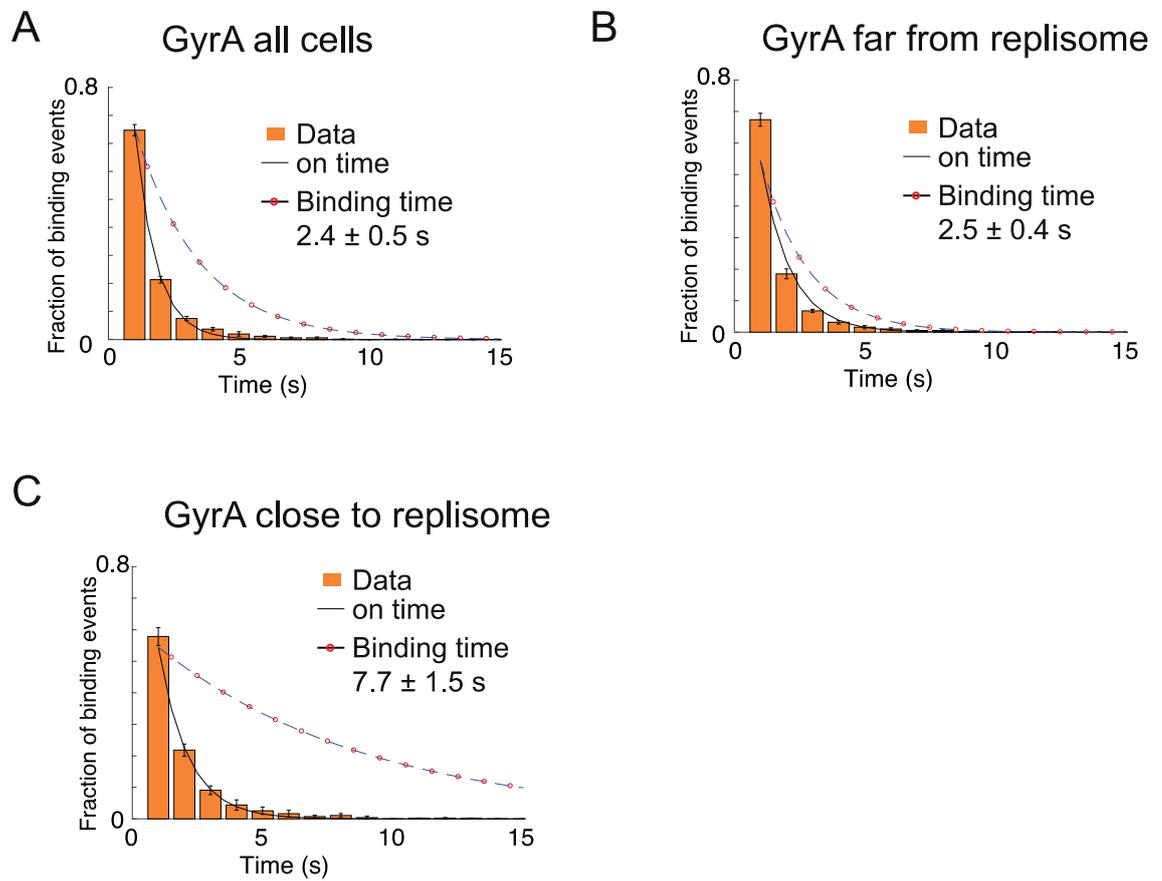

**Figure S4. Dwell times of GyrA.** On-time distributions for immobile GyrA-PAmCherry imaged with 1 s exposure times. Single exponential fits (solid lines) and photobleaching-corrected on time distributions (dashed circled lines). Photobleaching times were estimated by imaging, under the same conditions, cells with MukB-PAmCherry fusion, which has been shown to have a dwell time of ~50 s. Error bars shows S.E.M. of three experimental repeats.